# *Zero-calibration cVEP BCI using word prediction: a proof of concept*

*Federica Turi   Nathalie Gayraud   Maureen Clerc*

*Inria Sophia Antipolis-Mediterranée, Université Côte d'Azur - France*

*Introduction:*

Brain Computer Interfaces (BCIs) based on visual evoked potentials (VEP) [1] allow for spelling from a keyboard of flashing characters. Among VEP BCIs, code-modulated visual evoked potentials (c-VEPs) are designed for high-speed communication [2]. In c-VEPs, all characters flash simultaneously. In particular, each character flashes according to a predefined 63-bit binary sequence (m-sequence), circular-shifted by a different time lag. For a given character, the m-sequence evokes a VEP in the electroencephalogram (EEG) of the subject [3], which can be used as a template. This template is obtained during a calibration phase at the beginning of each session. Then, the system outputs the desired character after a predefined number of repetitions by estimating its time lag with respect to the template. Our work avoids the calibration phase, by extracting from the VEP relative lags between successive characters, and predicting the full word using a dictionary.

*Material, Methods and Results:*

Using the time-windowed EEG generated while the user is gazing at the first character, we compute the average response $X_a$ over N repetitions. Since the system has not been calibrated, the first character cannot be displayed. For the second character, we again compute the average response, and shift it by l·s time samples where s is the time lag between two consecutive characters. This produces L shifted averages $X_l$, l = {0, … , L-1}, where L is the number of characters on the keyboard. Using the lag $l$ = argmax$_l${corr($X_a$,$X_l$)} which produces the maximum correlation to the initial average response, we compute the relative position of this character with respect to the first. Finally, we generate all valid pairs of characters separated by l, and only retain those corresponding to the beginning of valid words within a dictionary. These word beginnings are displayed on the screen as feedback. We repeat this procedure for the following characters, until we are left with a single word (Fig.1). At that moment, we will have recovered the original letter, and the absolute position of $X_a$ can be thereafter used during the computation of the time lag. We conducted offline experiments using the database presented in [3], composed of 9 subjects, 2 sessions per subject, and 640 trials per session. The signals were pre-processed using a Butterworth filter between 1 and 15 Hz. Each experiment consisted of spelling a 3-letter word and was parameterized by the number of repetitions. We repeated the experiment 100 times by simulating the spelling of 3-letter words that we randomly selected among 1014 3-letter English words. We compared our results to a calibrated experiment (Fig.1b and 1c), where we used N repetitions of three characters to compute an average absolute response $X_a$, and performed the same pre-processing as in [3].

*Discussion:*

Our zero-calibration method achieves good accuracy, even with only 8 repetitions. In comparison, the experiments preceded by calibration reach a good accuracy after 12 repetitions of the m-code flashes. On Fig. 1b we distinguish two groups of subjects: in green, those that perform well, reaching on average an accuracy that exceeds 75% after 12 repetitions; in red, those whose performance does not produce a accuracy higher than 50%. This trend is also seen in the results of [3]. We keep the same color coding on Fig. 1c. While some subjects reach accuracy values equal to 100% after the calibration, others perform poorly, even compared to the zero-calibration method.

*Significance:*

Zero-calibration BCIs are widely research as their use is more natural. We have shown that a word prediction-based zero-calibration method in c-VEP BCIs can be efficient. Since this method relies on the correct detection of relative time lags, online experiments will be conducted to further determine the efficiency of our method.

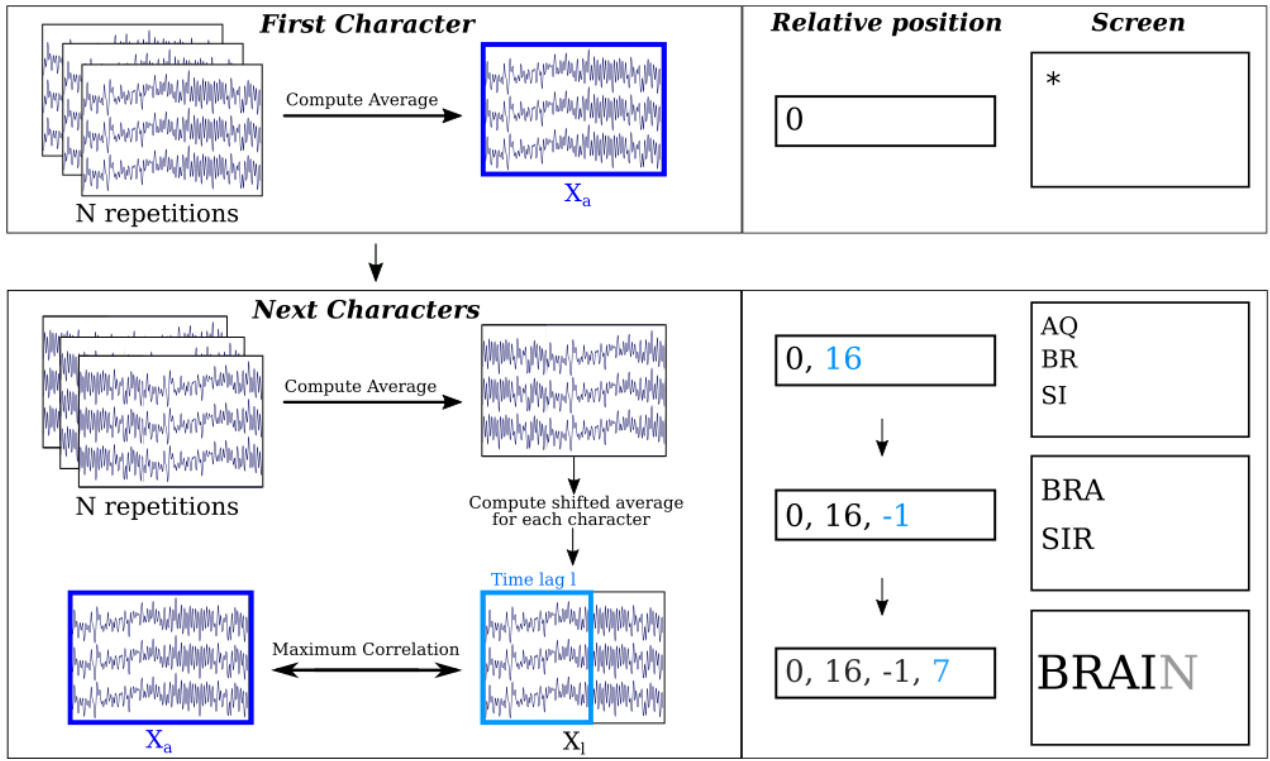

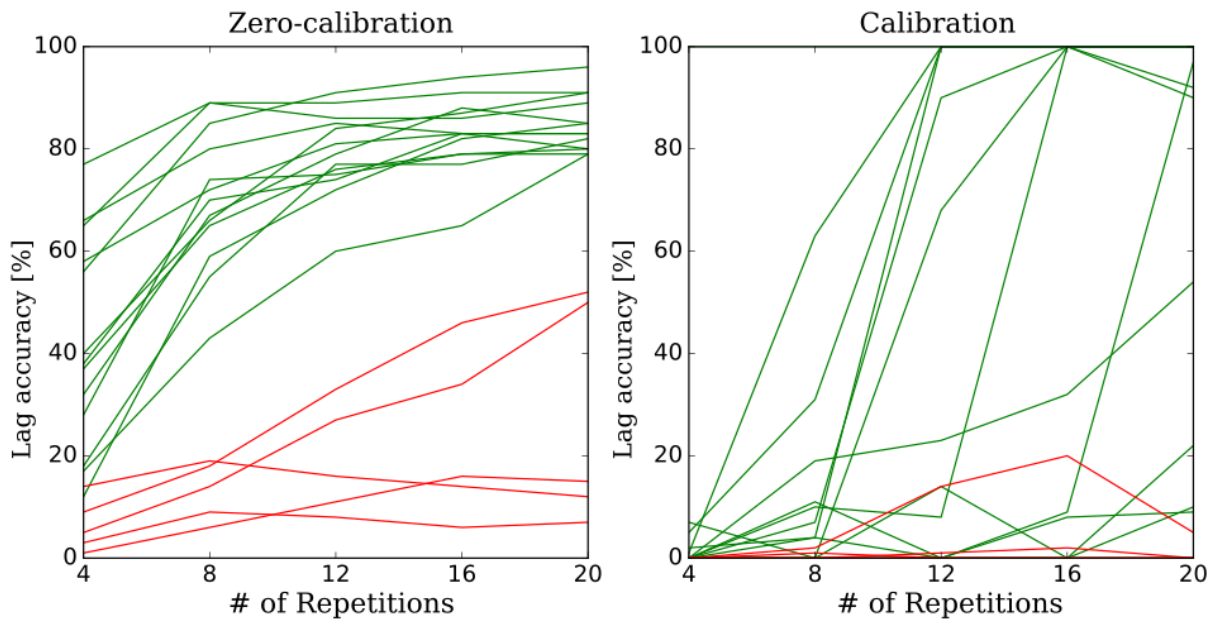

Figure 1. (a) Outline of the method: an example of spelling the word BRAIN up to the 4th letter. (b),(c) Average lag accuracy over all experiments for each subject and session, showing how many times each method was able to recover all the correct lags within a single word. The correct lag recovery is crucial to the performance of our method.